\documentclass[12pt,preprint]{aastex}
\shorttitle{Vela pulsar parallax} \shortauthors{P. A. Caraveo et
al.}
\begin{document}

\title{The distance to the Vela pulsar gauged with HST parallax
observations.\footnote{Based on observations with  the NASA/ESA
Hubble  Space Telescope, obtained at the Space Telescope Science
Institute, which is operated by AURA, Inc., under NASA contract
NAS 5-26555.}}

\author{P.A. Caraveo}
\affil{IFC-CNR, Via Bassini 15, I-20133 Milan}
\email{pat@ifctr.mi.cnr.it}
\author{A. De Luca }
\affil{IFC-CNR, Via Bassini 15, I-20133 Milan}
\author{R.P. Mignani}
\affil{ESO, Karl Schwarzschild Str.2, D8574O Garching b.
M\"unchen}
\author{G.F.Bignami}
\affil{ASI, Via Liegi 26, I-00198 Rome}

\begin{abstract}
The   distance  to  the   Vela  pulsar   (PSR  B0833$-$45)   has
been traditionally assumed to be 500 pc. Although affected by a
significant uncertainty, this value  stuck to both the pulsar
and  the SNR.  In an effort to  obtain a model  free distance
measurement, we  have applied high  resolution  astrometry  to
the  pulsar $V  \sim  23.6$  optical counterpart.   Using a  set
of  five HST/WFPC2  observations,  we have obtained the first
optical measurement of the annual  parallax of the Vela pulsar.
The parallax turns out to be $3.4 \pm 0.7$ mas, implying a
distance of $294^{-50}_{+76}$  pc, i.e.  a value significantly
lower than  previously believed.  This  affects the  estimate of
the pulsar absolute  luminosity  and  of   its  emission
efficiency  at  various wavelengths and confirms the
exceptionally  high value  of the  $N_e$ towards  the Vela
pulsar.   Finally, the  complete parallax  data base allows  for
a  better measurement  of  the Vela  pulsar proper  motion
($\mu_{\alpha}       cos(\delta)=-37.2\pm1.2$      mas yr$^{-1}$;
$\mu_{\delta}=28.2\pm1.3$  mas  yr$^{-1}$  after correcting  for
the peculiar motion of  the Sun) which, at the parallax distance,
implies a transverse velocity of $\approx$ 65 km sec$^{-1}$ .
Moreover, the proper motion position angle appears specially well
aligned with the axis of symmetry of the X-ray nebula as seen by
Chandra. Such an alignment allows to assess the space velocity of
the Vela pulsar to be 81 km sec$^{-1}$.
\end{abstract}


\section{Introduction}

Assessing  distances  to  Isolated  Neutron  Stars (INSs)  is  a
very challenging task which has  been pursued using different
techniques in different  regions  of  the  electromagnetic
spectrum.   Distances to pulsars with no glitching activity can be
obtained through radio timing techniques \citep{bel}.  However,
only millisecond radio pulsars allow for  positional accuracies
high enough  to be used for  parallax measurements.
\cite{toscano99} count 6 such cases in their list of 12 pulsars,
the distances to which have been determined through parallax.
For  the remaining 6 objects, which are  all classical  pulsars
with periods of  few hundreds msec, a  VLBI approach was  used,
requiring a suitable nearby calibrator.  The difficulties of the
VLBI technique, including the need to account for changing
ionospheric conditions, are apparent from the significant
revisions already published  for two of the six parallax values.
The  parallax of PSR B0919+06 went from 0.31 $\pm$ 0.14 mas
\citep{fom99} to  0.83 $\pm$ 0.14 mas \citep{chat01}, while for
PSR B0950+08 the parallax went from 7.9 $\pm$  0.8 mas
\citep{gwinn86} to 3.6 $\pm$ 0.3 mas \citep{brisken00}.  \\ Even
if limited to twelve objects, e.g. $<$1\%    of   the pulsar
family   \citep{cam00}, determining model-independent distances
of nearby pulsars is a rewarding exercise. As summarized  by
\cite{campbell96} and \cite{toscano99}, this allows to  trace the
local interstellar medium.   A distance value, coupled with the
pulsar dispersion measure, yields the electron density along the
line of sight, to be compared with the model of the galactic
$N_e$ distribution \citep{taco93}. Such a  model is  used to
derive the distances to all of  the remaining pulsars ($>$99\% of
the population). Moreover, a distance  value transforms the
pulsar proper motion into a firm transverse velocity, to be
compared with the average pulsar 3-D velocities  obtained by
\cite{lylo94}  and \cite{coche98} on larger samples. \\  X-ray
astronomy provides hints to  the distance of the  score  of
pulsars  detected  so  far \citep{bt97}  through  the
measurement  of   the  absorption of their  soft   X-ray
emission. Unfortunately,  distances derived from  X-ray
absorption  are  as uncertain as those derived by the dispersion
measure. In general, the radio distances  are greater than the
X-ray  ones.   \\  With  the detection of pulsars in the optical
(see e.g.  Caraveo 2000), distance measurements  became possible
using  classic optical astrometry techniques, based on parallax
measurements. Measuring tiny parallactic displacements is never
easy,  and the  task   can become really challenging when  the
targets  are intrinsically faint, like isolated neutron stars. It
requires high angular resolution coupled with high sensitivity,
rendering HST the instrument of choice to measure proper motions
and parallaxes of faint objects.  The astrometric capabilities of
HST were used to obtain  model free measurements of the distance
to Geminga \citep{car96} and RXJ 1856$-$3754 \citep{wal01}, two
radio quiet INSs for  which radio  astronomy could not  provide
any input.
\\ In this paper we address the
Vela pulsar (PSR B0833$-$45) which provides a compelling case of
nearby INS with a highly disputed value of the  distance and a
relatively bright optical counterpart ($V  \sim  23.6$). The
value  of 500  pc, tentatively  obtained  after the original
pulsar discovery  (e.g.  Milne  1968; Prentice  \&  ter Haar
1969), has  been assumed as  a reference for  both the pulsar and
its surrounding  SNR.   As  such,  it  is still  quoted  in
radio  pulsar catalogues \footnote{see                     e.g.
http://www.atnf.csiro.au/Research/pulsar/psr/archive/
\\      or http://pulsar.ucolick.org/cog/pulsars/catalog/},   in
spite   of  the doubts  raised by  several independent
investigations carried  out at different wavelengths, targeting
both the pulsar and the  SNR.  On the basis of their analysis of
the ROSAT data, \citet{psz96} placed the Vela pulsar  at 285 $\pm$
120 pc,  while \citet{pav01a},  using Chandra data  fitted by  a
two component  model, obtain a  distance of $210  \pm  20$ pc.
On  the  other  hand,  studying the  Vela  remnant \citet{cha99}
and \citet{bms99} find a distance of $250 \pm 30$ pc and
$\approx$  280 pc,  respectively.   A different  view  is
proposed  by \citet{gva01} who,  on the basis of the  pulsar
scintillation velocity and on  its very uncertain interpretation,
sees no reasons  to revise the canonical 500 pc distance. \\ We
started our observations aimed at the measurement of  the  Vela
pulsar parallax  in  1997, during  HST observing Cycle  6, but we
had to  wait until Cycle 8  to complete our program.   Meanwhile,
our  data have  also been used to  reassess the proper motion  of
the Vela  pulsar \citep{del00a} and to  quantify the overall
reliability of our astrometric approach  \citep{del00b}.  In the
following we shall report the analysis of our HST data leading to
the measurement of  the  Vela  pulsar  annual parallax  and proper
motion. The impact  of our result on the current understanding of
the Vela pulsar  is  also   discussed. While HST  was collecting
the appropriate set  of images, VLBI in the southern hemisphere
came into existence  and Vela  was one of the targets.  The
preliminary radio results \citep{leg00} will be compared to our
optical ones.

\section{The observations}

\begin{figure}
\plotone{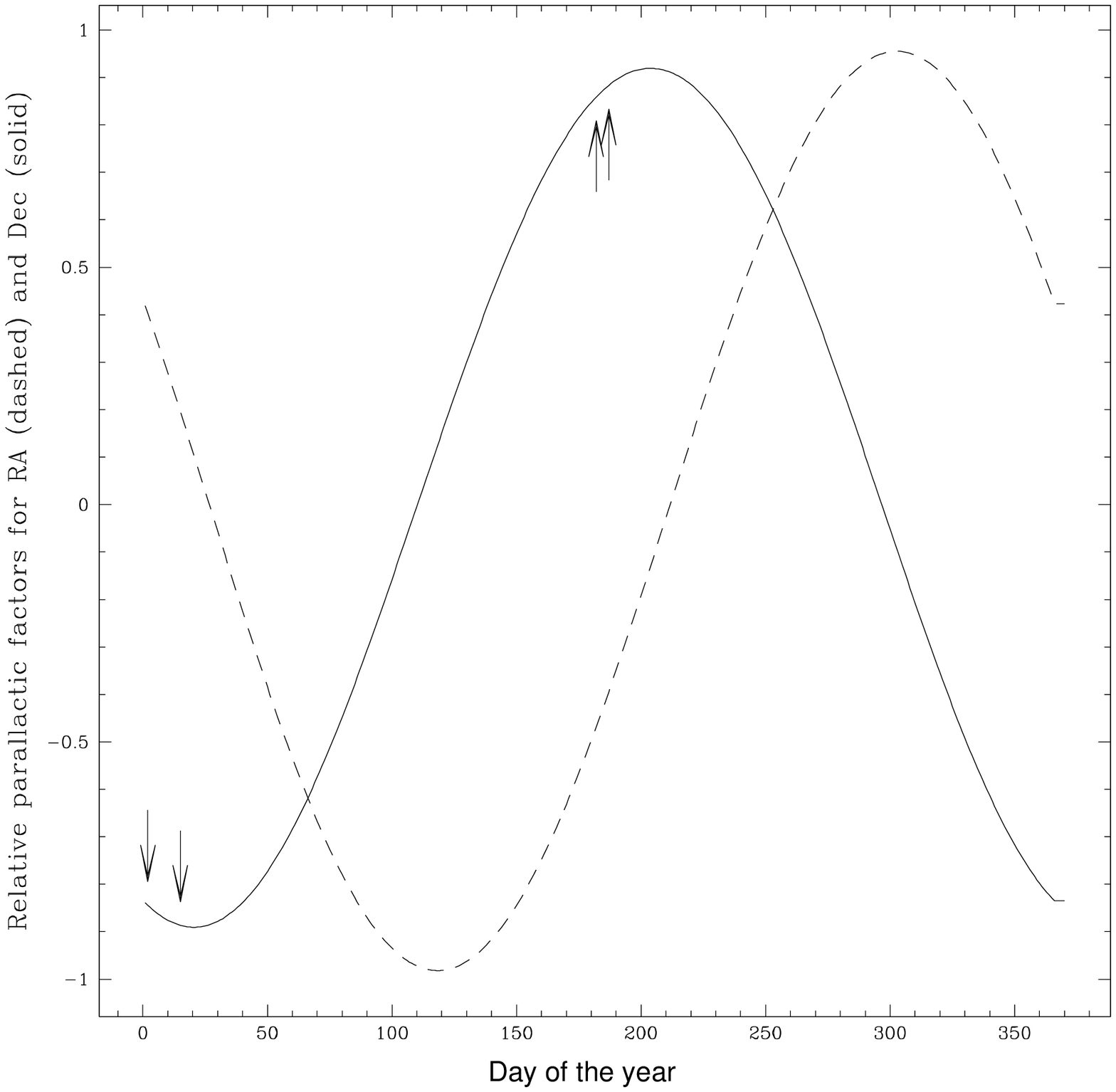} \caption{Relative parallactic  factors in right
ascension (dashed)
  and declination  (continuous) computed  for the  Vela pulsar
position. The arrows mark the periods of the year corresponding
to the observations listed in Table ~\ref{tbl1}. \label{fig1}}
\end{figure}

The  measurement of  the annual  parallax  of a  star through
optical astrometry techniques requires a set of at least three
observations of the field, preferably  taken at the epochs of the
maximum parallactic elongation.  For the Vela pulsar
($\alpha_{J2000}=08^{h}35^{m}20''.6$,
$\delta_{J2000}=-45^{\circ}10'35''.1$),   the    epochs   of
maximum parallactic elongation coincide with days  118 and 303 of
the year for right ascension,  and days 20  and 204 for
declination,  with relative parallax factors (see e.g. Murray,
1983) at the maximum elongation of  $\approx0.97$ and
$\approx0.90$, respectively (see  Figure   ~\ref{fig1}).
Although  the relative parallactic factor at maximum elongation
is somewhat larger in right ascension, observations at  one of
the  corresponding epochs turned  out  to be  difficult  to
schedule,  due  to  their very tiny visibility  window.  For
this  reason, the  observations of Vela were scheduled  close to
the  epochs  of  the   maximum parallactic displacements in
declination.  \\  Our program was originally approved for  HST
Cycle  6  but,  unfortunately, only two  observations of  the
planned triplet were  executed. The whole program had  to go
through a new approval  cycle and was rescheduled and
successfully completed in Cycle 8.  We thus obtained a  total of
five observations  of the field with the WFPC2 between June  1997
and July 2000.  The complete journal of the observations is
summarized in Table ~\ref{tbl1}. At each epoch, two exposures of
the field were acquired with  the WFPC2 ``V'' filter $F555W$
($\lambda=5252$\AA,$\Delta\lambda=1225$\AA) and  with similar
integration times.   In order to  maximize the angular
resolution, in all cases the pulsar optical counterpart was
centered on the Planetary Camera chip (PC) of the WFPC2 (pixel
size of 0.045 arcsec).

%\notetoeditor{table 1 should be placed here}
\section{Data reduction and analysis}
\label{datared}

\begin{figure}
\plotone{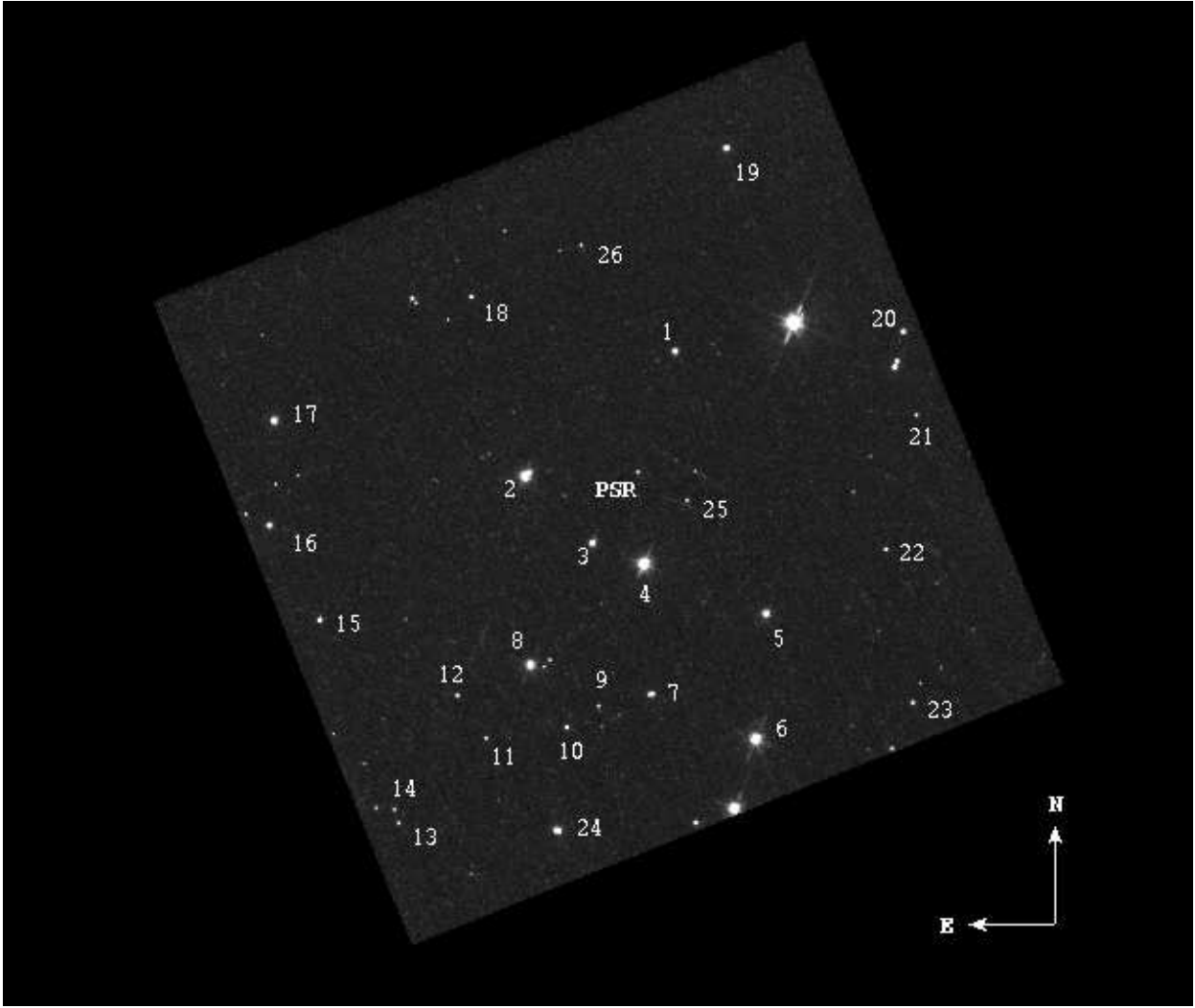} \caption{Image  of the  field of  the Vela
pulsar taken  with the Planetary Camera  of the WFPC2 and  filter
$555W$. The  pulsar and the reference stars used for astrometry
are labelled. \label{fig2}}\end{figure}

\begin{figure}
\plotone{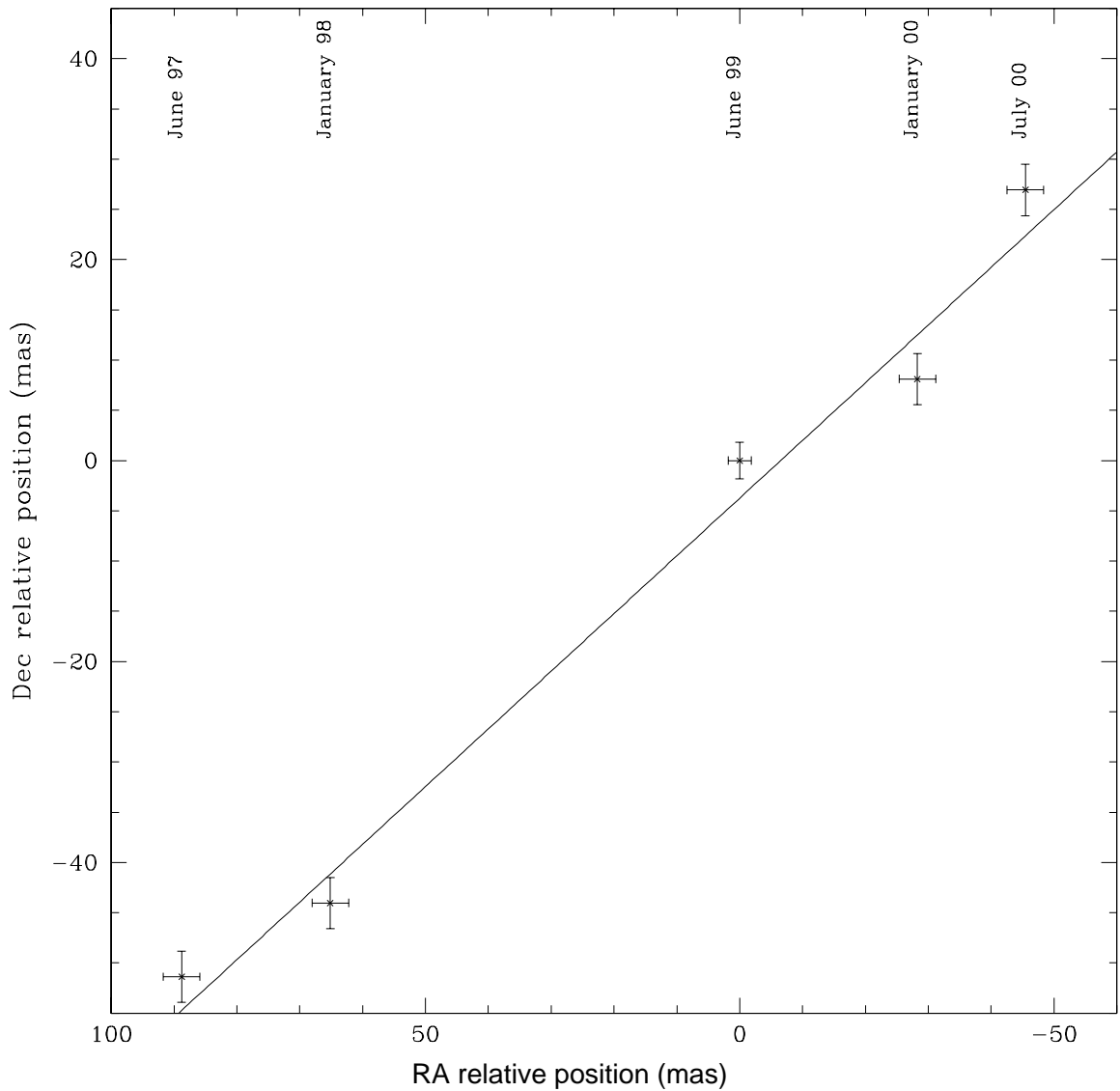} \caption{Relative  positions of  the pulsar
obtained  at  the five epochs.  North to  the top, East to the
left.   The error bars account for the overall  errors in the
pulsar positioning.   For the reference epoch (1999  June 30),
the  error bars (0.04 pixel,  $\approx1.8$ mas) reflect only  the
centering uncertainties  while for the  other epochs
($0.06\div0.07$  pixel, $\approx  2.5\div2.9$ mas)  we  have
accounted also for  the errors arising  from the frame
registrations.  The solid line corresponds to the best fit proper
motion \citep{del00b}. \label{fig3}}
\end{figure}

The data  reduction was been performed using  the IRAF/STSDAS
package. After the standard pipeline  processing of the frames
(debiasing, dark subtraction, flatfielding), which was  performed
using the most recent reference  files  and tables,  each couple
of  coaligned images  was co-added for  a first filtering  from
cosmic ray hits.   Residual hits were later rejected using
specific cosmic  ray subtraction algorithms in IRAF.   Figure
~\ref{fig2}  shows  the  resulting  image for  July
2000. \\
% with the pulsar position marked  by an arrow.
The  cleaned  images have  then  been used  for  the  definition
of  a relative  reference frame  to  be used  as  a starting
point for  our astrometric  procedure.  Since  all the
observations have  been taken with different telescope roll
angles and with small relative offsets, the  definition of  a
relative  reference frame  must rely  on  a very accurate image
superposition.    Following  the   approach  applied successfully
in previous astrometric works (e.g.  Caraveo et al. 1996;
Caraveo  and Mignani  1999;  De Luca  et  al.  2000a; Mignani et
al. 2000a,  Mignani et al.   2000b), this is done by computing a
linear coordinate   transformation  (i.e. accounting  for   2
independent translation factors, 2  scale factors and a rotation
angle) between a set  of reference objects.  \\  The selection of
the  reference grid objects  is critical.   They  must  be
point-like,  present  in  all observations (but not too close to
the field edges), bright  enough to  allow for  an accurate
positioning (but not saturated).  A set of 26 such objects
(labeled in Figure
~\ref{fig2}) was selected in the reference frame  of the PC.  \\
Since the shape of the PC Point Spread Function (PSF)  is known to
be position-dependent \citep{k95}, we did not use simulated PSFs
for source  fitting. On the other hand, the number of good
reference stars was not sufficient to  compute a template PSF
directly from the images. Thus, the reference object coordinates
were computed by fitting a  2-D  gaussian function  to  their
intensity profiles,  using optimized centering  areas.  This
yielded positional uncertainties of  the order of $0.01\div0.05$
pixel, depending on the objects' brightness  and position  on the
chip.   Special care  was devoted  to  characterizing  the
errors  involved  in  the centroid determination. Following
\citet{del00b},   we addressed   both statistical errors (i.e.
due to each object's S/N) and systematic ones (due, e.g., to the
telescope jitter, defocussing of the camera, charge transfer  in
the  CCD, background  fluctuations, etc.). Moreover, we checked
that our results were not biased by the algorithm used for the
fitting. The  coordinates were corrected  for  the ``34$^{th}$
row error'' \citep{ak99}  and for the  significant, well-known,
instrument geometrical distortion using  the most recent mapping
of  the PC field of view  \citep{cw00}.  The  centroids  of  the
Vela pulsar  optical counterpart  were  obtained  in  the  same
way, yielding errors  ranging between 0.02 and 0.04  pixel.   \\
Having secured a reference  grid,  we registered all the  frames
on the June 1999 one,  taken as a reference and previously
aligned along right ascension and declination according to  the
telescope  roll angle.  The rms of the residuals  on the
reference object coordinates  were $<0.05$ pixel in  right
ascension  and $<0.04$ pixel in  declination. The overall
accuracy  of the  frame registration, accounting  for errors in
the  centroid determination as well as in the geometric
distortions mapping, and the accuracy of the fit, has been
discussed in detail  by \cite{del00b}.
\\ To  ensure  that  our procedure was  not affected  by any
displacement  of our  26 reference objects  (due  either  to
proper  motions  or  parallaxes),  we  have repeated the  frame
registration 26  times, i.e. each  time excluding one of the
objects from the  fit. In addition, to exclude any possible bias
due to  the arbitrary  choice of  the reference  frame,  we have
repeated the whole procedure cycling  it over the five epochs. In
all cases, we obtained statistically undistinguishable results.
We  are thus confident  that our procedure  is correct  and  free
of  systematics. Last,  we have  applied  the  coordinate
transformations  to  the positions  of the Vela  pulsar; the
resulting relative positions are
shown in Figure ~\ref{fig3}. \\
If fitted with a simple proper motion, all the points in Figure 3
are seen to deviate from the straight line. Their residuals wrt
the proper motion fit, however, are not randomly distributed.
Rather, they follow the trend expected for an object affected
also by parallactic displacement.

\section{Analysis of the pulsar displacements}
\label{results} 
\begin{figure}
\plotone{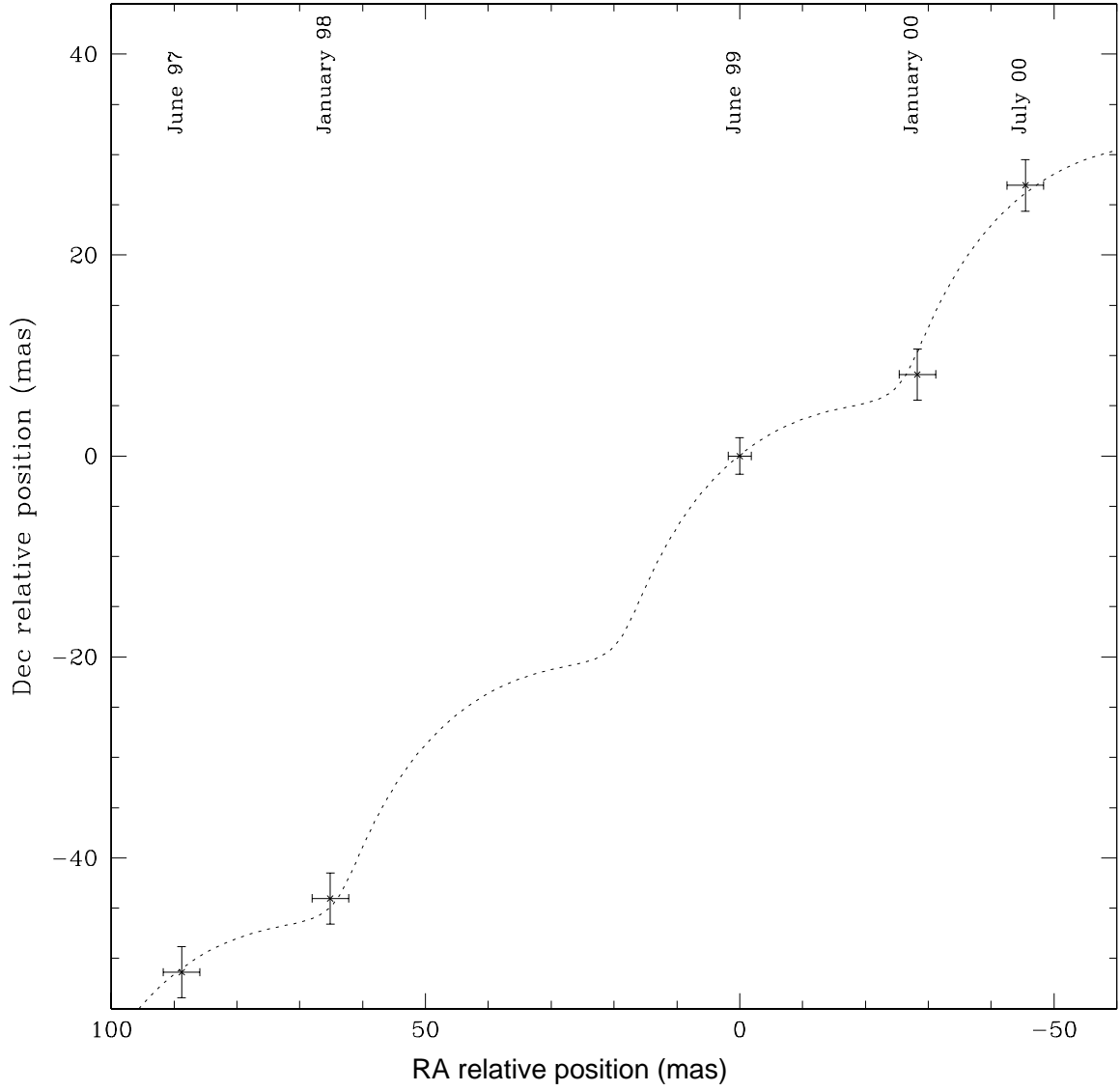} \caption{Same points as in Figure ~\ref{fig3}.
The dashed line has been derived using our best fitting parallax
and proper motion values.  \label{fig4}}
\end{figure}

The  geocentric  right  ascension  and
declination  $(\alpha_G  (t), \delta_G (t))$ of the pulsar at a
given epoch $t$ can be expressed in the form

\begin{equation} \left\{ \begin{array}{ll}
\label{eqs}
   \alpha_G (t)  =  \alpha_B (t_0) + \mu_{\alpha} cos(\delta) (t-t_0) +
\pi
P_{\alpha}(t) \\
   \delta_G(t)  =  \delta_B(t_0) +\mu_{\delta} (t-t_0) + \pi P_{\delta}
(t)
\end{array}
\right. \end{equation}

\noindent
where  $(\alpha_B  (t_0),  \delta_B   (t_0)  )$  are  the
barycentric coordinates at  a reference time  $t_0$,
$(\mu_{\alpha},\mu_{\delta})$ are  the right  ascension  and
declination  components  of the  proper motion,  $\pi$ is the
annual parallax  and $(P_{\alpha}(t),P_{\delta} (t))$ are the
parallactic factors shown in Figure ~\ref{fig1}.  A system of
equations like (~\ref{eqs}) can be written for each of the five
epochs corresponding to our observations.   \\ Setting
$t_0$=1999, June 30 as the reference epoch, one can obtain five
pairs of equations relating the  observed coordinates  of  the
pulsar to  the  5 unknowns  $\pi$, $\mu_{\alpha}$, $\mu_{\delta}$,
$\alpha_B      (t_0)$     and $\delta_B(t_0)$. A least  squares
fit yields for the  parallax and the proper  motion  the
following  values: $\pi=3.4$  mas, $\mu_{\alpha}
cos(\delta)=-45.0$  mas yr$^{-1}$, $\mu_{\delta}=25.8$  mas
yr$^{-1}$, corresponding to a reduced $\chi^{2}$  of 0.11 (5
degrees of freedom).
\\To  evaluate the  uncertainties  and the  confidence  levels on  the
fitted  parameters, we ran  a MonteCarlo  simulation for a
theoretical source featuring proper motion and parallax values
equal to our best fit ones. Each  synthetic data  set  was
obtained  by perturbating  a coordinate set  representing the
expected  source geocentric positions at  the  epochs  of  our
observations. The experimental  error  (per coordinate)  for each
data  point  in each  simulated data  set  was computed from a
normal distribution with a standard deviation equal to the overall
uncertainty (per coordinate) affecting the pulsar positioning
estimated in sect. ~\ref{datared}  (i.e. $\leq$2 mas per
coordinate in 1999,  June 30;  $2.5\div 2.9$  mas  per
coordinate  in the  remaining
epochs). \\
After $10^{5}$ simulations, we estimate the 1 ${\sigma}$ error
bar, to be attached to the parallax value, at 0.7 mas, while the
uncertainties on the proper motion are 1.1 mas in ${\alpha}$ and 1
mas in ${\delta}$. Thus, our best evaluation of the Vela pulsar
displacements is as follows:
% \\ $\pi=3.4\pm0.7$  mas ,\\
%$\mu_{\alpha}cos(\delta)=-45.0\pm1.1$ mas  yr$^{-1}$,\\
%$\mu_{\delta}=25.8\pm1.0$ mas yr$^{-1}$, corresponding to a P.A.
%$299.8^{\circ}\pm1.2^{\circ}$. \\

\begin{center}
$\pi=3.4\pm0.7 \hspace{2mm}$~mas \\
$\mu_{\alpha}cos(\delta)=-45.0\pm1.1$ \hspace{2mm} mas
\hspace{1mm}
yr$^{-1}$\\
$\mu_{\delta}=25.8\pm1.0$ \hspace{2mm} mas \hspace{1mm} yr$^{-1}$\\
\end{center}

corresponding to a position angle of
$299.8^{\circ}\pm1.2^{\circ}$.
\\The Vela path in the sky, as predicted from
the best fitting proper motion and parallax, is plotted in Figure
~\ref{fig4}, together with the measured pulsar positions. The
agreement between the expected positions and the measured ones is
remarkably good.

\subsection{The proper motion}
Our   best   fit   to   the  pulsar   proper   motion improves
the result  of \cite{del00b}  by confirming  the overall value
but by further reducing its associated errors.  Since we measure
the Vela proper  motion with  respect to  a set  of reference
stars in the  field, our value is sensitive to  the peculiar
motion of the Sun in the Local Standard of Rest. This induces an
apparent annual displacement of the pulsar in the anti-Apex
direction, which should be corrected for in  order to evaluate
the pulsar  motion with respect to the  structures located  in
its  immediate surroundings  (such  as the synchrotron nebula
detected in X-rays). A word of caution is required, since the
correction depends critically  on a reliable  definition of the
LSR. Following the analysis  of \citet{db98}, based on Hypparcos
data,  we   assumed  a  peculiar  motion  of   the  Sun  of
$13.4\pm0.8$    km/s     in    the    direction
$l=28\pm3^{\circ}$, $b=32\pm2^{\circ}$. The  components of the
proper motion of  the Vela pulsar    with    respect   to the
LSR   become    $\mu_{\alpha} cos(\delta)=-37.2\pm1.2$ mas
yr$^{-1}$, $\mu_{\delta}=28.2\pm1.3$ mas yr$^{-1}$,  for a
resulting  position angle  of $307.2\pm1.6^{\circ}$. We note
that    a    very    similar    result    ($\mu_{\alpha}
cos(\delta)=-38.3\pm1.2$ mas  yr$^{-1}$,
$\mu_{\delta}=27.4\pm1.3$ mas yr$^{-1}$,
P.A.=$305.6\pm1.6^{\circ}$)   is   obtained   using   the
correction for the Sun peculiar motion adopted by Bailes et
al.(1989).

\subsection{The annual parallax}
The value  of the pulsar annual  parallax which best fits  our
data is
  $3.4\pm0.7$  mas.    The  probability  of
  obtaining such a value in  absence of any parallactic displacement is
  of  the order  $10^{-5}$.   From  our parallax  value we  can
  derive the distance to the Vela pulsar, which turns out to be:

$$ 294^{-50}_{+76} \hspace{2mm} pc$$
\\
We can now compare our results  on the Vela proper motion and
parallax with  the  preliminary radio  ones  obtained  by
\citet{leg00},  e.g., $\mu_{\alpha}cos(\delta) = -49.8 \pm 0.2$
mas yr$^{-1}$, $\mu_{\delta} = 30.5 \pm  0.1$ mas yr$^{-1}$ and
$\pi = 3.13 \pm 0.33$  mas. While his parallax  value is similar
to ours, his proper  motion values are  not compatible with those
obtained  using the  HST both  for the study of "pure",
parallax-free, proper motion  \citep{del00b} and for our parallax
fit. However, no  details are available on the calibrator used,
nor  on  the   ionospheric  correction  approach,  nor on the
correction for galactic  rotation (on top of the Sun peculiar
motion), so that we cannot  assess the  accuracy of  the  radio
data analysis. Since  the optical approach  we have used is free
from all  such systematics, we conclude that  our distance
determination  based, as it is,  on direct measurement, with well
known error determination procedure,  is to be taken  as the most
reliable  estimate to  the true  distance  of this celestial
object.

\section{Discussions}

After  Geminga \citep{car96} and  RXJ1856-3754 \citep{wal01},
this is the third measurement  of the optical parallax of  an
isolated neutron star, to be  added to the list of a  dozen radio
parallaxes summarized by \cite{toscano99}.   At variance with the
two  previous cases, that had no firm distance estimates, the new
value for the distance of the Vela pulsar  is significantly
smaller than  the traditionally accepted one,  confirming the
earlier  claims by  \citet{psz96}, \citet{bms99}, \citet{cha99}
and \citet{pav01a}.  \\  While adding an important piece of
information  to  the   distribution  of  electrons  in  the  local
environment, a distance of  $\leq$300 pc has several implications
for the  Vela pulsar  physics, as  well as  for its kinematics.
In what follows we shall address each of these points.

\subsection{Local Interstellar Medium}

\cite{toscano99} used  the 12 radio pulsars with  a measured
parallax to  map the local  interstellar medium  and deduce
electron densities along their different lines of sight. The Vela
pulsar was added to the sample  on  the  basis  of  the distance
of  250  pc,  suggested  by \cite{cha99} for the  SNR. Coupling
the Vela DM  (67 cm$^{-3}$) with such a distance,
\cite{toscano99} computed an $N_e$ of  0.270 cm$^{-3}$, to be
compared  with   the  average  density   of  0.02  cm$^{-3}$ of
the \cite{taco93} model for the  local region.  Indeed, the $N_e$
towards Vela is  the maximum in their  sample and it  is 5 to 10
times higher than the values found for  four other pulsars in the
3$^{rd}$ galactic quadrant.   These  pulsars already show  $N_e$
values  systematically higher  than a  comparable number  of
objects in  the first  galactic quadrant.   Our distance value,
while  slightly  lowering the  $N_e$ value to  0.23 cm$^{-3}$,
confirms the extremely high electron density towards Vela,
possibly pointing towards the existence of ionised clouds in the
Vela region direction, at the  edge of the so called $\beta$ CMa
tunnel. This problem has been recently addressed by \citet{cha00}
who, doing the "astronephography" of the region (3-D mapping of
interstellar clouds),  found   three distinct absorption systems
with different  velocities, consistent  with an enhanced density
of ionized material towards PSR B0833-45.

\subsection{Multiwavelength emission}

For twenty years, up to 1993, the Crab  and Vela pulsars were the
only neutron stars detected  throughout the entire electromagnetic
spectrum, from radio to optical to  high-energy $\gamma$-rays.
Remarkably similar in high energy  $\gamma$-rays, the behavior of
the  two pulsars is very different    at all   other
wavelengths   (see e.g.  Thompson. 2001). Furthermore,  our
current understanding  of   the pulsar    emission mechanisms at
wavelengths other than radio  has   been biased by the
phenomenology  of  Crab and Vela,  since  theories have been
shaped to account for the multiwavelength behavior of  these two
objects.  The significant downsizing of the  Vela pulsar
luminosity  (by a factor of $\approx3 \pm 1$) presented here,
implies some revision on the current view of   pulsars
multiwavelength  behavior.  However,   lowering  the distance to
the  Vela   pulsar does not   have the same  impact in  the
various spectral domains.  \\  Vela has  a very  special place in
the high energy $\gamma$-ray sky where it is by  far the
brightest source, outshining the Crab  by a  factor  of
$\approx$4.  Its radiation   is totally pulsed \citep{gok94} and
its spectral shape varies throughout the  light curve.  Of
course,  to  convert the  measured flux  into a luminosity one
needs to know the  pulsar distance,  together with its beaming
factor.  While the  $\gamma$-ray community never disputed  the
traditional 500 pc  value for the   Vela distance, the beaming
solid angle remains unknown, ranging from the size of a neutron
star polar cap  to $4\pi$ for isotropic emission  (see Thompson
et al. 1999 and references therein).  For simplicity, a 1 sr
conical beam is generally assumed, yielding a beaming factor of
$1/4\pi$. With such a beaming value, the high energy luminosity
of the Vela pulsar is now $7 \times 10^{33}$ erg sec$^{-1}$,
comparable to that of the much older PSR B1055$-$52 and 10 times
lower that that of PSR B1706$-$44, a pulsar remarkably  similar
to Vela in its  $P$, $\dot P$ and overall energetics.   Moreover,
the efficiency with which Vela  converts its rotational energy
loss  into $\gamma$-rays becomes 0.001,  similar to that of the
much younger Crab and  20 times smaller  than that of PSR
B1706$-$44.   Not surprisingly, the Vela pulsar  is now well
below any best  fitting line  correlating the $\gamma$-ray
luminosity to pulsar parameters such as the number  of
accelerated particles  \citep{tho99} or the  value of the open
field  line voltage \citep{tho01}. This will require a critical
re-examination of the position of Vela in the $\gamma$-ray
emitting  pulsar   family.    Vela may    become an
underluminous  pulsar, an apparent  paradox considering its
brightness in the  $\gamma$-ray sky.  \\  In the  X-ray domain,
the  situation is different.  The X-ray emission phenomenology
already prompted a number of authors (Page  et al. 1996; Pavlov
et al.   2001a; Helfand et  al. 2001) to opt  for a   distance
smaller than the ``canonical''    one. Indeed, for a distance of
500 pc, the  soft X-ray flux, most probably originated by the hot
surface of the neutron  star, pointed towards an emitting area
($R^{\infty}=3\div4$ km)   far too small to be compatible with
the whole neutron  star  surface \citep{ofz93}.  On the  other
hand, a  very small emitting hot spot   is obviously incompatible
with the  shallow pulsation seen  at  these energies.    Reducing
the pulsar distance to  half its ``canonical'' value,
\citet{pav01a} find that for a pure black-body emission, a radius
of the emitting area of just    $1.9\div2.4$ km could explain
the  flux observed by Chandra. Introducing a modified black body
with  a pure H atmosphere eases the emitting  area problem,
allowing for  the whole  surface of   a 13 km radius neutron star
at  210 pc to   emit at a temperate  significantly smaller than
that obtained in the   pure black body  case.   Our new
independent distance determination now  freezes one of  the
parameters of the model, allowing for a better  determination of
both the neutron star   radius and its   temperature.
\\The parallax distance fixes the overall Vela X-ray luminosity
(0.2-8 keV) to be about 5.5 10$^{32}$ erg/sec, divided between a
thermal component, dominating at low energies, and a non thermal
one, emerging only at higher energies. If compared to the
available rotational energy loss, such an X-ray emission accounts
for 8 10$^{-5}$ of the pulsar reservoir, significantly less than
the average ratio of $\approx$  10$^{-3}$ found for the majority
of the young X-ray emitting pulsars \citep{bt97}. Moreover, such
an already low conversion efficiency shrinks to 10$^{-6}$ if one
considers only the non-thermal component, confirming the severe
underluminosity  of PSR 0833-45 in the X-ray domain. The
interpretation of such an underluminosity is, however, complicated
by the composite nature of the Vela emission. It is certainly
dominated by thermal emission  for E $\leq$ 1.8  keV, with a
weaker magnetospheric component emerging only  at E $\geq$ 1.8
keV \citet{pav01a}. Because of such a composite nature, the Vela
pulsar  X-ray emission  cannot  be readily compared  to other
purely magnetospheric cases (e.g the Crab and PSR B0540$-$69).
\\The X-ray data can also be used to further map the local ISM.
The comparison between distance and Hydrogen column density,
$N_H$, responsible of the X-ray absorption, provides  an
assessment of the  reliability  of   this
parameter as a   distance indicator for nearby   sources.   \\
In  the optical domain,  the reduction in the  pulsar luminosity
has important implications.  Here the    emission is certainly
magnetospheric   - as witnessed by its  double-peaked light curve
\citep{wal77,man78, gou98} and by its flat  $3600\div8000$ \AA ~
spectrum \citep{nas97,mc01}. However, the revised value of the
B-to-U optical luminosity $L_{BU}$ $\approx$ 5.5 10$^{28}$ erg
sec$^{-1}$  is now quite low with respect to the prediction of
the classical \citet{ps83} model, based on synchrotron emission
at the light cylinder. Such an underluminosity could neither be
ascribed to the spectral shape of the optical emission, which is
remarkably similar to the Crab one (Mignani and Caraveo, 2001),
nor to the shape of the light curve (see e.g. Gouiffes, 1998 ),
which covers a broader phase interval than the Crab one,
resulting, if anything, in an increase of the optical output.
\\The behaviour of the Vela pulsar in the optical domain should be
considered in the light of its ``transition'' position  between the
group of the young energetic pulsars, characterized by pure
magnetospheric emission and the middle-aged ones characterized by
composite spectra and lower emission efficiency (see e.g.
Caraveo, 1998; Mignani and Caraveo, 2001). \\In the X-ray domain
Vela behaves as a middle-aged object (see e.g. Pavlov et al,
2001a) while in the optical and gamma-ray energy ranges its
phenomenology is reminiscent of that of the younger pulsars but
with a low emission efficiency. The multiwavelength luminosities
of the Vela pulsar, stemming from the newly determined distance,
will help to discriminate the emission mechanisms at work in the
different energy domains and to assess their evolution throughout
the life of the pulsars.

\subsection{Proper motion}
\label{pm}

The parallax fit yields also, as  a by-product, the best proper
motion value   obtained so far at  optical  wavelengths,
corresponding to a transverse velocity of $\approx$ 65 km
sec$^{-1}$.  The space direction of  the Vela proper motion   has
been correlated with   the axis  of symmetry of the Chandra X-ray
structure  by a   number of authors (Mignani et al.  2000b,
Helfand et  al.   2001  and Pavlov et al. 2001b). Since  the
X-ray  nebula axis  of symmetry should  trace the pulsar
rotational axis, an alignment between such axis of symmetry and
the pulsar proper motion would have important bearings of the
kick mechanism responsible for the  neutron star ejection.
\citet{lai01}, noting that proper motion-axis of  symmetry
alignment  seems to  be present also  for   the  Crab  pulsar
\citep{cm99}, considered different spin-kick alignment
mechanisms.  Also in view of the pulsars' low velocities, they
concluded that both natal kicks, be they hydrodynamic or
neutrino-magnetic, as well as postnatal ones, such as the
Harrison and Tademaru (1975)
electromagnetic kick, could account for the alignments.\\
Here, we reassess the case for the Vela proper motion X-ray jet
alignment by comparing our "corrected" proper motion position
angle (p.a.) of $307.2^{\circ}\pm1.6^{\circ}$ with the value of
$307^{\circ} \pm 2^{\circ}$ computed by \citet{pav01b} for the
axis of  symmetry of the structure seen by Chandra. Assuming an
uniform probability distribution for the two vectors in space, we
estimate the chance occurrence probability to be 1.2 $10^{-3}$,
thus strengthening the case for an alignment between the  Vela
rotation  axis  and its proper motion. Figure 5 shows an update
of Figure 2 of \citet{mcd00}, with our corrected proper  motion
vector superimposed  to the high resolution Chandra  image of the
Vela nebula. Taking advantage of such an alignment we can use the
angle of 53.2$^{\circ}$, computed by Helfand et al (2001) between
the axis of the X-ray torus and the line of sight, to convert our
measured transverse velocity into the pulsar 3-D velocity. Thus,
the space velocity of the Vela pulsar turns out to be 81 km/sec.
While hardly compatible with the \citet{lylo94} mean value of 400
km sec$^{-1}$, our value is also in the low side  of the low
velocity component of the bimodal fit by \citet{coche98},
classifying Vela as a rather slow moving pulsar.\\

\section{Conclusions}

We have presented the first direct measurement of  the distance
to the Vela pulsar at optical wavelengths.  Our value is
significantly smaller than the canonical  one, implying a
downgrading of the  luminosity of the Vela pulsar,   positioning
Vela among  the underluminous  pulsars. High energy $\gamma$-ray
astronomy is the most  affected by our result since it pertains
to the brightest, hence  most conspicuous, source in the
$\gamma$-ray sky.  The   optical domain  is also  seriously
affected by our overall luminosity reduction, while  the X-ray
side is only marginally altered.  The distance determination also
confirms the anomalously   high  density of    electrons  towards
the  Vela pulsar, pointing to the  presence of ionized clouds  in
the  local interstellar medium.  \\ As  a byproduct of the
parallax fit, we have  obtained a very accurate proper motion
value, which  appears nicely aligned along the  axis of symmetry
of the X-ray nebula,  hence the pulsar rotation axis. Moreover,
our  proper motion,   together with   our  distance
determination, fixes the transverse velocity of  Vela to 65 km
sec$^{-1}$. Using the inclination of the X-ray torus, as measured
by Helfand et al (2001), we can convert our transverse velocity
into the 3-D one, yielding a value of 81 km sec$^{-1}$. For a
pulsar, this is certainly a low velocity.

\section{Acknowledgements}

This work was supported by the Italian Space Agency (ASI). \\We
thank the referee, David Chernoff, for his suggestions.

\clearpage

\clearpage

\clearpage

\clearpage

\clearpage

\begin{figure}
\plotone{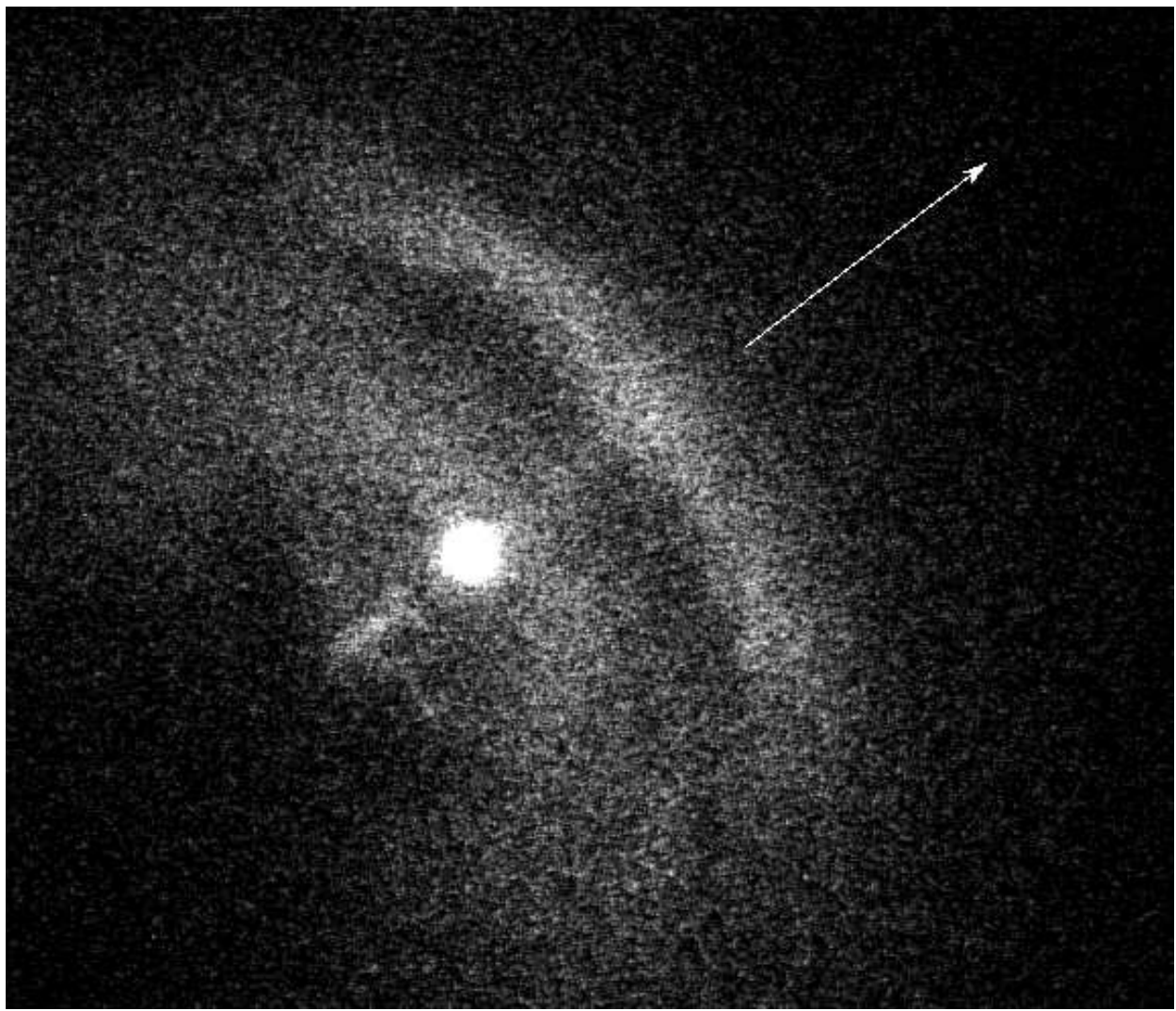} \caption{X ray image of the Vela pulsar and
synchrotron nebula taken with the Chandra High Resolution Camera
(North to the top, East to the left). The image has been smoothed
with a low-pass filter in order
 to highlight the large scale structures.
The arrow indicates the direction of the pulsar proper motion
corrected for the peculiar motion of the Sun as discussed in
Section ~\ref{pm}. }
\end{figure}

\clearpage

\begin{table}
\begin{tabular}{c|c|c|c|c|c} \tableline \tableline
%\label{dataset}
Obs. ID & Date &$P_{\alpha}$ & $P_{\delta}$ &  N.of exp. & Exposure(s) \\
\tableline \tableline
  1 & 1997 June 30    & -0.467 & 0.850 & 2 & 1300 \\
  2 & 1998 January 2  & 0.399 & -0.852 & 2 & 1000 \\
  3 & 1999 June 30    & -0.478 & 0.852 & 2 & 1000 \\
  4 & 2000 January 15 & 0.196 & -0.887 & 2 & 1300 \\
  5 & 2000 July 5     & -0.394 & 0.883 & 2 & 1300 \\ \tableline \tableline
\end{tabular}
\caption{Summary   of  the   HST/WFPC2  observations   used for
the measurement of the Vela pulsar parallax. In all cases the
observations were taken with the same  instrument set-up, i.e.,
through the $F555W$ filter and with  the target positioned at
the  center of the Planetary Camera. For  each observation, the
columns give the  sequence number, the observing epoch and the
corresponding parallactic factor in right ascension and
declination ($P_{\alpha}$,$P_{\delta}$), the  number of repeated
exposures and exposure  time in seconds.  Observations \#1 to \#4
are the ones used by  \citet{del00a} and \citet{del00b} to
reassess the Vela pulsar proper motion. \label{tbl1}}
\end{table}

\end{document}